\documentclass[12pt,a4paper]{article}
\usepackage[cp866]{inputenc}
\usepackage{amsmath}
\usepackage{graphicx}
\usepackage{epsfig}
\newcommand{\be}{\begin{equation}}
\newcommand{\ee}{\end{equation}}
\newcommand{\ben}{\begin{equation*}}
\newcommand{\een}{\end{equation*}}
\newcommand{\bea}{\begin{eqnarray}}
\newcommand{\eea}{\end{eqnarray}}
\newcommand{\ar}{\begin{array}}
\newcommand{\arn}{\end{array}}

\newcommand{\vk}{\vec{k}}
\newcommand{\vks}{\vec{k}^{\;2}}
\newcommand{\q}{\vec{q}}
\newcommand{\qs}{\vec{q}^{\;2}}

\newcommand{\vl}{\vec{l}}
\newcommand{\vls}{\vec{l}^{\;2}}

\newcommand{\x}{\vec{r}}
\newcommand{\xs}{\vec{r}^{\;2}}
\newcommand{\xp}{\vec{r}^{\;\prime}}
\newcommand{\xps}{\vec{r}^{\;\prime\; 2}}

\newcommand{\ec}{\vec{e}^{\;*}}

\def\pnot{\mbox{${\not{\hbox{\kern-3.0pt$p$}}}$}}
\def\qnot{\mbox{${\not{\hbox{\kern-2.0pt$q$}}}$}}
\def\enot{\mbox{${\not{\hbox{\kern-2.0pt$e$}}}$}}
\def\knot{\mbox{${\not{\hbox{\kern-2.0pt$k$}}}$}}

\def\fun#1#2{\lower3.6pt\vbox{\baselineskip0pt\lineskip.9pt\ialign
{$\mathsurround=0pt#1\hfil##\hfil$\crcr#2\crcr\sim\crcr}}}

\begin{document}
\sloppy                              
\renewcommand{\baselinestretch}{1.0} 

\begin{titlepage}

\hskip 11cm \vbox{ \hbox{Budker INP 2014-2}  }
\vskip 3cm
\begin{center}
{\bf  Impact factors for Reggeon-gluon transition in $N=4$ SYM with large number of colours}
\end{center}

\vskip 0.5cm

\centerline{V.S.~Fadin
$^{a\,\dag}$, R.~Fiore$^{b\,\ddag}$}

\vskip .6cm

\centerline{\sl $^{a}$
Budker Institute of Nuclear Physics of SD RAS, 630090 Novosibirsk
Russia}
\centerline{\sl and Novosibirsk State University, 630090 Novosibirsk, Russia}
\centerline{\sl $^{b}$ Dipartimento di Fisica, Universit\`a della Calabria,}
\centerline{\sl and Istituto Nazionale di Fisica Nucleare, Gruppo collegato di Cosenza,}
\centerline{\sl Arcavacata di Rende, I-87036 Cosenza, Italy}

\vskip 2cm

\begin{abstract}
We calculate impact  factors for Reggeon-gluon transition  in supersymmetric Yang-Mills theory with four supercharges at large number of colours $N_c$.  In the next-to-leading order  impact factors are not uniquely defined and must accord with BFKL kernels and energy scales. We obtain  the impact factor corresponding to the kernel and the energy evolution parameter,  which is invariant under M\"{o}bius transformation in  momentum space,   and show that it is also M\"{o}bius  invariant up to terms taken into account in the BDS ansatz.
\end{abstract}


\vfill \hrule \vskip.3cm \noindent $^{\ast}${\it Work supported 
in part by the Ministry of Education and Science of Russian Federation,
in part by  RFBR,  grant 13-02-01023, and in part by Ministero Italiano dell'Istruzione,
dell'Universit\`a e della Ricerca.} \vfill $
\begin{array}{ll} ^{\dag}\mbox{{\it e-mail address:}} &
\mbox{fadin@inp.nsk.su}\\
^{\ddag}\mbox{{\it e-mail address:}} &
\mbox{roberto.fiore@cs.infn.it}\\
\end{array}
$
\end{titlepage}

\vfill \eject

\section{Introduction}

In  the BFKL approach ~\cite{BFKL}, impact factors appear as an integral part.  Scattering amplitudes of high energy processes are given in this approach by convolutions of Green
functions of interacting Reggeized gluons with the impact factors of scattered  particles,
therefore the notion of these impact factors is well known. Less known are the impact factors
for Reggeon-particle (in particular Reggeon-gluon)  transitions, where for Reggeon here and in the following we mean Reggeized gluon. They appeared  firstly
\cite{Fadin:2002et} in 
the proof of the  multi-Regge form of QCD amplitudes. An  idea  of this form  is  the basis of the BFKL approach. It  appeared \cite{{Lipatov:1976zz},  {BFKL}}  from results of fixed order calculations.  Later  it was proved in the leading logarithmic approximation (LLA) \cite{Balitskii:1979} with the use of the $s$-channel unitarity. The proof of the  multi-Regge form  in the next-to-leading approximation  (NLA)  is based also on the  $s$-channel unitarity \cite{Fadin:2006bj}.  Compatibility of the   unitarity with the multi-Regge form  leads to  bootstrap relations connecting  discontinuities of the amplitudes with products of their real parts and gluon trajectories.  It turns out \cite{{Fadin:2002et},{Fadin:2006bj}} that   the fulfillment of an infinite set of these relations guarantees the multi--Regge form of scattering  amplitudes. On the other hand,  all bootstrap relations  are fulfilled if several conditions imposed on the Reggeon vertices and the trajectory (bootstrap conditions) hold true.   The most complicated condition, which includes the impact factors for Reggeon-gluon transition, was
proved  recently,  both in QCD \cite{Kozlov:2011zza} - \cite{Kozlov:2012zza} and in its supersymmetric generalisation \cite{Kozlov:2013zza}.

Recently,  the  impact factors for Reggeon-gluon  transition were used for the calculation of the high-energy behavior of the remainder function to the BDS ansatz \cite{Bern:2005iz} for multi-particle   amplitudes with maximal helicity violation (MHV amplitudes) in Yang--Mills theory,  with maximally extended supersymmetry (N=4 SYM) in the limit of large number of colours. It was shown \cite{Bartels:2008ce} that in the so called Mandelstam kinematical region  the BDS amplitude $M^{BDS}_{2\rightarrow 4}$ should be multiplied by the factor containing the contribution of the Mandelstam cut, and this contribution for the 6-point scattering amplitude was found in the leading logarithmic approximation (LLA) \cite{Bartels:2008sc} and in the next-to-leading one (NLA)  \cite{Lipatov:2010qg}-\cite{Fadin:2011we}.

In the BFKL approach this contribution  is given by the convolution of the Green function of two interacting Reggeons  with  the impact factors for Reggeon-gluon transition.
In the NLA the remainder function was calculated \cite {Fadin:2011we}
assuming the existence of conformal invariant (in momentum space) representations of the modified (i.e. with the subtracted gluon trajectory depending on the total momentum transfer) BFKL kernel for the adjoint representation of the gauge group  and  impact factors for  Reggeon-gluon transition. Later it was shown \cite{Fadin:2013hpa} that indeed the modified BFKL kernel has the  conformal invariant representation. As for the impact factors, actually not the impact factors themselves, but the   convolution of two impact factors (which was called for brevity also impact factor) was used. Moreover, the convolution used in  Ref.~\cite {Fadin:2011we} was not calculated in the framework of the BFKL approach, but was extracted \cite{Lipatov:2010ad}  from the two-loop 6-point remainder function obtained in  Ref.~\cite{Goncharov:2010jf} by simplification of the results of  Refs.~\cite{DelDuca:2009au} and \cite{DelDuca:2010zg}.  In turn, in the derivation of these results it was supposed that the remainder function appears as expectation value of Wilson loops in N=4 SYM.  All this makes
the direct calculation of the impact factors for Reggeon-gluon transition in the BFKL framework  and the investigation of their properties  very important.

In this paper  we calculate  the impact  factors for the Reggeon-gluon transition  in the next-to-leading order  (NLO) for $N=4$ SYM with a large number of colours, i.e. in the planar approximation.  As it is well known, the NLO impact factors are not uniquely defined (they are scheme dependent) and must accord with BFKL kernels and energy scales (energy evolution parameters). Our aim is to find the impact factor which corresponds to  the conformal invariant kernel found in Ref.~\cite{Fadin:2013hpa}  and to the energy scale used in Ref.~\cite {Fadin:2011we}.  Just this impact  factor, with  the deduction of terms contained  in the BDS ansatz, is expected to be invariant under M\"{o}bius transformation in momentum space,  according to the conjecture (not yet proved) about the dual conformal  invariance of the remainder function. We reach this aim  starting from the impact factor in the ``bootstrap scheme", which was  found in Refs.~\cite{Kozlov:2011zza} -- \cite{Kozlov:2013zza} in Yang-Mills theories with  any  number of fermions and scalars in arbitrary representations of the  gauge group.  Using these results and the known relation between the bootstrap scheme and the scheme defined in Ref.~\cite{Fadin:2006bj},
which is called  standard scheme,  we obtain  the  impact factor for  $N=4$ SYM in the last  scheme.  In this scheme, however,  neither  the BFKL kernel, nor the energy evolution parameter  are   M\"{o}bius invariant. Therefore, to obtain the  impact factor, which is supposed to be M\"{o}bius invariant (after subtraction of terms included in the BDS ansatz), one has to transform  the standard impact factor so as to accord it  with   the M\"{o}bius invariant kernel found in Ref.~\cite{Fadin:2013hpa} and with the M\"{o}bius invariant evolution parameter. If the arguments for the dual conformal invariance of the remainder function are correct, the result should be M\"{o}bius invariant, up to terms kept in the BDS ansatz. Below we demonstrate that it is the case.

  The paper is organized as follows. In
the next section we calculate in the planar approximation the impact factor in $N=4$ SYM  in the bootstrap scheme. In Section 3 this impact factor is transformed  into the standard scheme. In Section 4 the   result obtained in Section 3 is transformed into the scheme with conformal kernel and energy evolution parameter (we call this scheme  M\"{o}bius scheme). Conclusions
are drawn in Section 5.


\section{The impact factor in  the bootstrap scheme}
In the Born approximation, with  the denotations and state normalizations  used in Refs.~\cite{Fadin:2006bj}-\cite{Kozlov:2013zza},  the impact factor for  the transition  of a Reggeon  $R$ with transverse (to the plane of initial momenta $p_A, p_B$) momentum $\q_1$  into a gluon $G$ with transverse momentum $\vk$ and polarization vector $e(k)$
in interaction with two Reggeized gluons ${\cal G}_1$ and ${\cal G}_2$ is written
as
\begin{equation}
\langle G  R_1|{\cal G}_{1}{\cal G}_{2}\rangle^{(B)}
= 2g^2\delta(\q_1-\vk-\x_1-\x_2) \left(T^{a}T^{b}\right)_{c_1c_2 }\; \ec\vec{C}_1.
\label{GRBorn}
\end{equation}
Here $g$ is the  coupling constant,  $T^i$ are the colour group generators, $\x_1$, $\x_2$ and $c_1$, $c_2$ are the  transverse momenta and colour indices  of the Reggeized gluons ${\cal G}_1$ and ${\cal G}_2$
correspondingly, $a$ and $b$ are the colour indices  of the Reggeon $R$
and the gluon $G$,  $\ec$ is the conjugated transverse part of  the  polarization vector $e(k)$ in the gauge $e(k)p_2=0$ with the  lightcone vector  $p_2$ close to the vector $p_B$, and
\begin{equation}
\vec{C}_1 \ = \ \q_1 - (\q_1-\x_1)\frac{\qs_1}{(\q_1-\x_1)^2}.
\label{vector C 1}
\end{equation}

In  N=4 SYM the NLO impact factor contains gluon, fermion and scalar contributions.  These contributions were found in Refs. \cite{Kozlov:2012zz} -\cite{Kozlov:2013zza} for Yang-Mills theories with  any  number of fermions and scalars in arbitrary representations of the  gauge group.

In general, the impact factors contain two parts with different colour structure. In the planar limit, which we are interested in,  only parts with the Born  colour structure remain.
They are given  by Eq.~(61) in Ref.~\cite{Kozlov:2011zza}, Eq.~(61) in  Ref.~\cite{Kozlov:2012zza}  and Eq.~(123) in Ref.~\cite{Kozlov:2013zza}
for fermions, gluons and scalars correspondingly.  Note however that these equations were derived using the dimensional  regularization, which differs from the dimensional reduction used in supersymmetric theories.  To take into account this difference we have to take the number $n_S$ of the scalar fields  equal to $6-2\epsilon$ (here and below $\epsilon =(D-4)/2, \;\; D$ being  the space-time dimension).
With account of this,  we obtain (details will be given elsewhere),
\begin{equation}
\langle G  R_1|{\cal G}_{1}{\cal G}_{2}\rangle \ = \
g^2\delta(\q_1-\vk-\x_1-\x_2) \left(T^{a}T^{b}\right)_{c_1c_2 }\; \ec\Bigg[2\vec{C}_1+ \bar{g}^2\vec{\Phi}_{G  R_1{*}}^{{\cal G}{\cal G}_{2}}\Bigg],
\label{GR}
\end{equation}
where
\[
\vec{\Phi}_{GR_1{*}}^{{\cal G}{\cal G}_{2}}=
\vec{C}_1\Biggl(\ln\left(\frac{(\q_1-\x_1)^2}
 {\vks}\right)\ln\left(\frac{\xs_2}{\vks} \right)
 +\ln\left(\frac{(\q_1-\x_1)^2\qs_1}{\vk^{\;4}}\right)\ln\left(\frac{\xs_1}{\qs_1} \right) -4\frac{(\vks)^\epsilon}{\epsilon^2}+6\zeta(2)\Biggr)
\]
\[
+\vec{C}_2
\Biggl(\ln\left(\frac{\vks}{\xs_2}
\right)\ln\left(\frac{(\q_1-\x_1)^2}{\xs_2}\right)
+\ln\left(\frac{\qs_2}{\qs_1}\right)\ln\left(\frac{\vks}{\qs_2}\right)\Biggr)
\]
\be
-2\Bigl[\vec{C}_1\times \Bigl(\Bigl[\vk\times \x_2\Bigr]I_{\vk, \x_2}-\Bigl[\q_1\times \x_1\bigr]I_{\q_1,  -\x_1}\Bigr)\Big] +2\Big[\vec{C}_2\times \Bigl(\Bigl[\vk\times \x_2\Bigr]I_{\vk, \x_2}+\Bigl[q_1\times \vk\bigr]I_{\q_1,  -\vk}\Bigr)\Big]~. \label{impact-star}
\ee
Here $\bar g^2=g^2\Gamma(1-\epsilon)/(4\pi)^{2+\epsilon}$ (note that in the expression \eqref{impact-star}  and in the following only terms not vanishing at $\epsilon\rightarrow 0$ should be  kept),
\begin{equation}
\vec{C}_2 \ = \ \q_1 - \vk\frac{\qs_1}{\vks},
\label{vector C 2}
\end{equation}
$\Gamma(x)$ is   the Euler gamma-function,  $\zeta(n)$ is the Riemann zeta-function ($\zeta(2)=\pi^2/6$), $\bigl[\vec a \times c \bigl[ \vec b\times \vec c\bigr]\bigr]$ is a double vector product, and
\begin{equation}
I_{\vec p, \vec q}=
\int_{0}^{1}\frac{dx}{(\vec p +x\vec q)^{2}}\ln\left(\frac{\vec p^{\;2}}
{x^2\vec q^{\;2}}\right), \;\; ~~~~~I_{\vec p,\vec q}=I_{-\vec p,-\vec q}=I_{\vec q, \vec p}=I_{\vec p,-\vec p
-\vec q}~. \label{I p q 1}
\end{equation}
Note that the expression \eqref{impact-star} is obtained after huge cancellations between gluon, fermion and scalar contributions. In particular, solely  due to  these cancellations only two vector structures ($\vec{C}_1$ and $\vec{C}_2$) remain; each of the contributions separately contains three independent vector structures.

As it was already mentioned, NLO corrections  are scheme dependent. The scheme used in the derivation of $\langle G  R_1|{\cal G}_{1}{\cal G}_{2}\rangle$ (given in Eqs.~(\ref{GR}  and  (\ref{impact-star})) was adjusted simplifying the verification of the bootstrap conditions (we call it bootstrap scheme). It is different from  the scheme defined in
Ref.~\cite{Fadin:2006bj} (in turn, we call it standard scheme).  The impact factors in these schemes are connected   by the transformation \cite{Kozlov:2012zz}
\begin{equation}
\langle G R_1|=\langle G R_1|_{s}- \langle G R_1|^{(B)}\widehat{\cal U}_{k},
\label{botstrap-standard transformation}
\end{equation}
where the subscript $s$ means the standard scheme and
the operator $\widehat{\cal U}_{k}$ is defined by the matrix elements
 \begin{equation}
 \langle {\cal G}_1'{\cal G}_2'|\widehat{\cal U}_{k}|{\cal G}_1{\cal G}_2\rangle =\frac{1}{2}
 \ln\Bigl(\frac{\vks}{(\x_1-\xp_1)^2}\Bigl)\langle {\cal
 G}_1'{\cal G}_2'|\widehat{\cal K}_r^B|{\cal G}_1{\cal G}_2\rangle.
 \label{operatorUk}
 \end{equation}
Here  $\widehat{\cal K}_r^B$ is the part of the LO BFKL kernel related to the real gluon production:
\begin{equation}
  \langle{\cal G}_1'{\cal G}_2'|\widehat{\cal K}_r^B|{\cal G}_1{\cal G}_2\rangle =\delta(\xp_1+\xp_2-\x_1-\x_2) \frac{g^2}{(2\pi)^{D-1}}T^i_{c_1c'_1}T^i_{c'_2c_2} \Big(\frac{\xs_1\xps_2+\xs_2\xps_1}{\vls}-\qs_2\Big),
\label{K born}
 \end{equation}
where $\vl=\x_1-\xp_1=\xp_2-\x_2$,\;\; $\q_2=\x_1+\x_2=\xp_1+\xp_2$.


\section{Transformation to  the standard  scheme}

From Eqs.~\eqref{GRBorn}--\eqref{GR} and \eqref{botstrap-standard transformation}--\eqref{K born} it follows that at large number $N_c$ of colours we can write
\[
\vec{\Phi}_{GR_1{s}}^{{\cal G}{\cal G}_{2}} \ = \    \vec{\Phi}_{GR_1{*}}^{{\cal G}{\cal G}_{2}}  + \vec{\cal I}_1,\;\;
\]
\begin{equation}
 \vec{\cal I}_1 =\int\frac{d\vl}{\Gamma(1-\epsilon)\pi^{1+\epsilon}} \vec{C}^{\;\prime}_1\frac{1}{\xps_1\xps_2}  \biggl(\frac{\xs_1\xps_2+\xs_2\xps_1}{\vls} -\qs_2\biggr)\ln\Big(\frac{\vks}{\vls}\Big),
\label{I 1}
\end{equation}
where
\begin{equation}
\vec{C}^{\;\prime}_1 \ = \  \q_{1}- \qs_1\frac{(\q_1-\xp_1)}{(\q_1-\xp_1)^{2}}.
\end{equation}
When  $\epsilon \rightarrow 0$,  the integral $\vec{\cal I}_1 $ is  infrared divergent at  $\vl =0$.  To calculate this integral, it is convenient to use the decomposition
\begin{equation}
\vec{C}^{\;\prime}_1 \ = \ \vec{C}_1 +\vec{\Delta}_1, \;\; \vec{\Delta}_1   \ = \ \qs_1\Big(\frac{(\q_1-\x_1)}{(\q_1-\x_1)^{2}}- \frac{(\q_1-\xp_1)}{(\q_1-\xp_1)^{2}}\Big).
\end{equation}
Then, the divergency will appear only  in the term with $\vec{C}_1$, which does not depend on  $\vl$ and can be taken outside of  the integral sign. After that, using the basic integrals
\[
\int\frac{d\vl}{\Gamma(1-\epsilon)\pi^{1+\epsilon}}\frac{1}{(\vec q -\vec l)^2(\vec p +\vec
l)^2}\ln \left(\frac{\vec l^{\:
2}}{\mu^2}\right)=\left((\vec q +\vec p)^2\right)^{\epsilon-1}\left[\frac{1}{\epsilon}\ln\left(\frac{\vec p^{\:
2}\vec q^{\:
2}}{\mu^4}\right)+\frac{1}{2}\ln^2\left(\frac{\vec p^{\:
2}}{\vec q^{\: 2}}\right) \right]+{\cal O(\epsilon)}~,
\]
\begin{equation}
\int\frac{d\vl}{\Gamma(1-\epsilon)\pi^{1+\epsilon}}\frac{1}{\vec l^{\: 2}(\vec q -\vec
l)^2}\ln \left(\frac{\vec l^{\:
2}}{\mu^2}\right)
= \left({\vec q^{\:
2}}\right)^{\epsilon-1}\left[-\frac{1}{\epsilon^2}+\zeta(2)
+ \frac{2}{\epsilon}\ln \left(\frac{\vec q^{\:
2}}{\mu^2}\right)\right]+{\cal O(\epsilon)}~,
\label{I2l}
\end{equation}
we obtain
\[
\int\frac{d\vl}{\Gamma(1-\epsilon)\pi^{1+\epsilon}} \frac{1}{\xps_1\xps_2}  \biggl(\frac{\xs_1\xps_2+\xs_2\xps_1}{\vls} -\qs_2\biggr)\ln\Big(\frac{\vks}{\vls}\Big)
\ = \ 2\frac{(\vks)^\epsilon}{\epsilon^2}-2\zeta(2) -\ln^2\big(\frac{(\x_1+\x_2)^2}{\vks}\big)
\]
\begin{equation}
- \ln\big(\frac{\xs_1\xs_2}{(\vks)^2}\big)\ln\big(\frac{\xs_1\xs_2}{(\x_1+\x_2)^4}\big) + \ln\big(\frac{\xs_1}{(\x_1+\x_2)^2}\big)\ln\big(\frac{\xs_2}{(\x_1+\x_2)^2}\big) .
\label{integral with C}
\end{equation}
The integral with $\vec{\Delta}_1$
is infrared finite  and can be calculated at $\epsilon=0$. It is convenient to calculte it  using
``helical" vector components $\pm$ instead of the Cartesian ones $x, y$
($\;a^{\pm} = a_x\pm ia_y$) and the  decomposition
\[
\frac{1}{\qs_1}  {\Delta}^+_1\frac{1}{\xps_1\xps_2}  \biggl(\frac{\xs_1\xps_2+\xs_2\xps_1}{\vls} -\qs_2\biggr)
\]
\[
=\frac{1}{(q_1-r_1)^-}\biggl[\frac{r_2^-}{k^-}\bigl(
\frac{1}{(r_1-l)^+}+\frac{1}{l^+}\bigr)\bigl(
\frac{1}{(r_2+l)^-}-\frac{1}{(q_1-r_1+l)^-}\bigr)
\]
\begin{equation}
+\frac{r_1^-}{q_1^-}\bigl(\frac{1}{l^+}
-\frac{1}{(r_2+l)^+}\bigr)\bigl(
\frac{1}{(r_1-l)^-}+\frac{1}{(q_1-r_1+l)^-}\bigr)\biggr].
\label{decomposition 1}
\end{equation}
Note that each term in this decomposition gives an ultraviolet divergent contribution to the integral \eqref{I 1} (of course, the total integral is ultraviolet convergent). Therefore, we introduce the ultraviolet cut-off $\Lambda \rightarrow \infty$.  Integrals with separate terms in the decomposition \eqref{decomposition 1} are calculated using the basic integral
\[
\int\frac{d \vec l}{\pi}\frac{1}{(a- 1)^+}\frac{1}{(b-1)^-}\ln\left(\frac{\vec l^{\;2}}{\mu^2}\right)
\theta(\Lambda^2-\vec l^{\;2})=
\frac12\ln\left(\frac{\Lambda^2}{(\vec a -\vec b)^{2}}\right)
\ln\left(\frac{\Lambda^2(\vec a -\vec b)^{2}}{\mu^4}\right)
\]
\begin{equation}
+\frac12\ln\left(\frac{(\vec a -\vec b)^{2}}{\vec b^{\;2}}\right)
\ln\left(\frac{(\vec a -\vec b)^{2}}{\vec a^{\;2}}\right)
+\frac{a^+b^--a^-b^+}{2}I_{\vec a,-\vec b }~, \label{int master}
\end{equation}
where $I_{\vec a,\vec b }$ is defined in Eq.~(\ref{I p q 1}). With the help of  this integral, one has
\[
\int\frac{d\vl}{\pi}\vec{\Delta}_1 \frac{1}{\xps_1\xps_2}  \biggl(\frac{\xs_1\xps_2+\xs_2\xps_1}{\vls} -\qs_2\biggr)\ln\Big(\frac{\vks}{\vls}\Big)
\]
\[
\ = \
\frac{1}{2}\vec{C}_1\Bigg(\ln\frac{\qs_2}{\xs_1}\ln\frac{\vk^{\;4}}{\xs_1\xs_2}+\ln\frac{(\q_1-\x_1)^2}{\vks}\ln\frac{\vk^{\;4}}{(\q_1-\x_1)^2\xs_2}\Bigg)
\]
\[
+\frac{1}{2}\Bigl(\vec{C}_1- \vec{C}_2\Bigr)\Bigg(\ln\frac{\qs_2}{\xs_2}\ln\frac{\vk^{\;4}}{\xs_1\xs_2}+\ln\frac{(\q_1-\x_1)^2}{\qs_1}\ln\frac{\vk^{\;4}}{\xs_1(\q_1-\x_1)^2} \Bigg)
\]
\begin{equation}
+\Bigl[\vec{C}_1\times[\vk\times \x_2]\Bigr]I_{\vk,\x_2}
+\Bigl[\vec{C}_2\times[\x_1\times \x_2]\Bigr]I_{\x_1,\x_2}
-\Bigl[\Bigl(\vec{C}_1-\vec{C}_2\Bigr)\times[\q_1\times \x_1]\Bigr]I_{\q_1,-\x_1}
~.
\label{integral with Delta}
\end{equation}
Using Eqs.~\eqref{integral with C} and \eqref{integral with Delta}  we obtain
\[
\vec{\cal I}_1  =\frac12 \vec{C}_1\Bigg[  \ln\left(\frac{\xs_2}{\vks}\right)\ln\left(\frac{\vk^{\;4}}{(\q_1-\x_1)^2\xs_2} \right)+\ln\left(\frac{\xs_1}{\vks}\right)\ln\left(\frac{\vks\qs_1}{(\q_1-\x_1)^2\xs_1} \right)
\]
\[
-\ln\left(\frac{\vks}{(\q_1-\x_1)^2}\right)\ln\left(\frac{\vks\qs_1}{(\q_1-\x_1)^4} \right) +4\frac{(\vks)^\epsilon}{\epsilon^2}-4\zeta(2)  \Bigg]
\]
\[
-\frac12\vec{C}_2\Bigg[
 \ln\left(\frac{\qs_2}{\xs_2}\right)\ln\left(\frac{\vk^{\;4}}{\xs_1\xs_2} \right)
+\ln\left(\frac{(\q_1-\x_1)^2}{\qs_1}\right)\ln\left(\frac{\vk^{\;4}}{\xs_1(\q_1-\x_1)^2} \right)\Bigg]
\]
\begin{equation}
+\Bigl[\vec{C}_1\times \Bigl(\Bigl[ \vk \times \x_2  \Bigr]
I_{\vk,\x_2}
-\Bigl[\q_1\times\x_1\Bigr]I_{\q_1,-\x_1}\Bigr)\Bigr]
+\Bigl[\vec{C}_2\times
\Bigl(\Bigl[\x_1\times\x_2\Bigr]I_{\x_1, \x_2}+\Bigl[\q_1 \times \x_1\Bigr]I_{\q_1,-\x_1}\Bigr)\Big]
~. \label{Integral I}
\end{equation}
The one-loop correction to  the  impact factor in the standard scheme is given by  Eqs.~\eqref{I 1},  \eqref{impact-star} and \eqref{Integral I} and reads
\[
\vec{\Phi}_{G  R_1{s}}^{{\cal G}_{1}{\cal G}_{2}}= \frac12\vec{C_1}\Biggl[ \ln\left(\frac{\qs_1}{\xs_1}\right)\ln\left(\frac{\vks\xs_1}{\q_1^{\;4}} \right)
 +\ln\left(\frac{(\q_1-\x_1)^2}{\vks}\right)\ln\left(\frac{\vk^{\;4}\xs_1}{(\q_1-\x_1)^4\qs_1} \right)
 \]
\[
 +\ln\left(\frac{\xs_2}{\vks}\right)\ln\left(\frac{(\q_1-\x_1)^2}{\xs_2} \right)-4\frac{(\vks)^\epsilon}{\epsilon^2}+8\zeta(2)\Biggr]
 +\Bigl[\vec{C_1}\times\Bigl( \Bigl[\q_1 \times \x_1\Bigr]I_{\q_1,-\x_1}-\Bigl[\vk \times r_2\Bigr]I_{\vk,\x_2}\Bigr)\Bigr]
\]
\[
+\frac12 \vec{C_2}
\Biggl[\ln\left(\frac{\qs_2}{\qs_1}\right)\ln\left(\frac{\xs_1\xs_2}{\q_2^{\;4}}\right)
+\ln\left(\frac{\xs_2}{(\q_1-\x_1)^2}\right)\ln\left(\frac{\xs_2\qs_1}{\xs_1(\q_1-\x_1)^2}\right)\Biggr]
\]
\begin{equation}
+\Bigr[\vec{C_2}
  \times\Bigl(\Bigl[\x_1\times \x_2\Bigr]I_{\x_1, \x_2}+\Bigl[\q_1\times \x_1\Bigr]I_{\q_1,-\x_1}+2\Bigl[\vk\times \x_2\Bigr]I_{\vk,\x_2}+2\Bigl[\q_1\times \vk\Bigr]I_{\q_1, -\vk}\Bigr)
  \Bigr] ~.\label{impact-standard}
\end{equation}
The correction \eqref{impact-standard}
fit the standard kernel \cite{Fadin:1998fv} and the energy scale $|\vk_1||\vk_2|$, where
$\vk_{1,2}$  are the transverse momenta of produced gluons in the two impact factors connected by the Green function of the two interacting Reggeons (BFKL ladder).


\section{The impact factor in the  M\"{o}bius scheme}

The impact factor in the  M\"{o}bius scheme means the impact factor for  Reggeon-gluon transition  which can be used  for the calculation of the remainder function with conformal invariant  kernel and energy evolution parameter.  Let us remind here that  the kernel used for  the calculation of the remainder function
\cite{Bartels:2008ce}-\cite{Fadin:2011we} (which is called modified kernel) is the BFKL kernel in $N=4$ SYM for the adjoint representation of the gauge group with the subtracted gluon trajectory depending on the total momentum transfer (the subtraction is made to avoid double counting of terms included in the BDS ansatz).

To obtain  the impact factor in the  M\"{o}bius scheme from the correction \eqref{impact-standard}  we have to perform two transformations,  to reconcile the impact factor  with the kernel and the energy scale.
As it was shown in Ref.~\cite{Fadin:2013hpa}, the conformal invariant $\hat{\cal K}_c$ and the standard $\hat{\cal K}_m$ forms of the modified kernel  are connected by the similarity transformation
\begin{equation}
\hat{\cal K}_c  = \hat{\cal K}_m -\frac14\left[\hat{\cal K}^B, \left[\ln\left(\hat{\vec{q}}_{1}^{\;2}\hat{\vec{q}}_{2}^{\;2}\right),
\hat{\cal{K}}^B\right]\right]\,,   \label{kernel transformtion}
\end{equation}
where  $\hat{\cal K}^B$  is the usual LO kernel and $\hat{\vec{q}}_{1,2}$ are the operators of  the Reggeon momenta. Note that in the commutator there is no difference between the usual and modifided kernels, so that $\hat{\cal K}^B$ is taken instead of  $\hat{\cal K}^B_m$.
The corresponding transformation for the impact factor is
\begin{equation}
 \langle G R_1|_{t} = \langle G R_1|_s -\frac14 \langle G R_1|^{(B)}\Bigg[\ln\Bigg(\hat{\vec{q}}^{\;2}_1\hat{\vec{q}}^{\;2}_2\Bigg), \hat{\cal K}^{(B)} \Bigg]\,,
 \label{transform to conform}
\end{equation}
where the subscript $t$ means transformed to fit the conformal kernel. For the NLO correction we obtain
\[
\vec{\Phi}_{GR_1{t}}^{{\cal G}{\cal G}_{2}} \ = \    \vec{\Phi}_{GR_1{s}}^{{\cal G}{\cal G}_{2}}  + \vec{\cal I}_2,\;\;
\]
\begin{equation}
 \vec{\cal I}_2 =\frac12\int\frac{d\vl}{\Gamma(1-\epsilon)\pi^{1+\epsilon}} \vec{C}^{\;\prime}_1\frac{1}{\xps_1\xps_2}  \biggl(\frac{\xs_1\xps_2+\xs_2\xps_1}{\vls} -\qs_2\biggr)
 \ln\Big(\frac{\xs_1\xs_2}{\xps_1\xps_2}\Big).
\label{I 2}
\end{equation}
This integral is infrared finite and can be calculated in two-dimensional space, with the help of the decomposition \eqref{decomposition 1}, the decomposition
\[
\frac{1}{\xps_1\xps_2}  \biggl(\frac{\xs_1\xps_2+\xs_2\xps_1}{\vls} -\qs_2\biggr)
\]
\begin{equation}
=-\left(\frac{1}{(r_1-l)^+}+\frac{1}{l^+}\right)\left(\frac{1}{(r_2+l)^-}-\frac{1}{l^-}\right)
-\left(\frac{1}{(r_1-l)^-}+\frac{1}{l^-}\right)\left(\frac{1}{(r_2+l)^+}-\frac{1}{l^+}\right)
\end{equation}
and the integral \eqref{int master}.
Using the result of integration,
\[
\vec{\cal I}_2  = \frac14 \vec{C}_2 \Bigg[
 \ln\left(\frac{\qs_2}{\xs_2}\right)\ln\left(\frac{\q^{\;4}_2}{\xs_1\xs_2} \right) +\ln\left(\frac{\qs_1}{\vks}\right)\ln\left(\frac{\xs_2}{\qs_2} \right)
  +\ln\left(\frac{(\q_1-\x_1)^2}{\qs_1}\right)\ln\left(\frac{\vks\qs_1}{\xs_1\xs_2} \right)\Bigg]
\]
\[
-\frac14\vec{C}_1\Bigg[\ln\left(\frac{\xs_1}{\xs_2}\right)\ln\left(\frac{\vks\xs_1}{\qs_1\xs_2} \right)
+\ln\left(\frac{\vks\qs_1}{\xs_1\xs_2}\right)\ln\left(\frac{(\q_1-\x_1)^4}{\vks\qs_1} \right)\Bigg]
\]
\[
+\Big[\vec{C}_1\times\Big(\Bigl[\vk\times\x_2\Bigr]I_{\vk,\x_2}-\Bigl[\q_1\times\x_1\Bigr]I_{\q_1,-\x_1}\Big)\Big]
\]
\begin{equation}
+\frac12\Big[\vec{C}_2\times\Bigl(\Bigl[\x_1\times \x_2\Bigr]I_{\x_1,\x_2}+\Bigl[\q_1\times\x_1\Bigr]I_{\q_1,-\x_1}-\Bigl[\vk\times\x_2\Bigr]I_{\vk,\x_2}-\Bigl[\q_1\times\vk\Bigr]I_{\q_1, -\vk}
   \Bigr)\Bigr] ~,\label{transform to quasi conform}
\end{equation}
we obtain the correction to the transformed impact factor:
\[
\vec{\Phi}_{G  R_1{t}}^{{\cal G}_{1}{\cal G}_{2}}=\vec{C}_1\Biggl[
\ln\left(\frac{\qs_2}{\qs_1}\right)\ln\left(\frac{\q_2^{\;4}(\q_1-\x_1)^4}{\qs_1\xs_2\vks\xs_1}\right)
-\ln\left(\frac{\qs_2(\q_1-\x_1)^2}
 {\qs_1\xs_2}\right)\ln\left(\frac{\qs_2(\q_1-\x_1)^2}{\vks\xs_1}\right)\
\]
\[
-\frac34\ln^2\left(\frac{\vks\xs_1} {\qs_1\xs_2}\right) - \ln^2\left(\frac{\qs_1}{\qs_2}\right)-2\frac{(\vks)^\epsilon}{\epsilon^2}+4\zeta(2)\Biggr]
\]
\[
+\frac14\vec{C}_2\ln\biggl(\frac{\qs_2(\q_1-\x_1)^2}{\qs_1\xs_2}\biggr)\ln\biggl(\frac{(\q_1-\x_1)^4\qs_1\vks\xs_1}{\x_2^{\;6}\q_2^{\;4}}\biggr)
\]
\begin{equation}
+\frac32 \Biggl[\vec{C}_2\times\Biggl(\Bigl[\x_1\times\x_2\Bigr]I_{\x_1,\x_2}+\Bigl[\q_1\times\x_1\Bigr]I_{\q_1,-\x_1}+\Bigl[\vk\times\x_2\Bigr]I_{\vk,\x_2}+\Bigl[\q_1\times\vk\Bigr]I_{\q_1, -\vk}
\Biggr)   \Biggr] . \label{impact-quasi conform}
\end{equation}
The M\"{o}bius  invariant kernel was used for the calculation of the NLO remainder function
in Ref.~\cite{Fadin:2011we} with the M\"{o}bius invariant convolution of the NLO BFKL impact
factor (which was called for brevity simply impact factor) obtained
in Ref.~\cite{Lipatov:2010qg} from direct two-loop calculations and with the energy scale
$s_0$ chosen in such a way that the ratio (energy evolution  parameter) $s/s_0=s\q_2^{\;2}/\sqrt{\q_1^{\;2}\q_3^{\;2}\vks_1\vks_2}$  is M\"{o}bius invariant.  This
energy scale differs from the energy scale  used  in the correction (\ref{impact-standard}) of the  impact factor  (see, for instance,  Ref.~\cite{Fadin:2006bj} ) which is equal to $|\vk_1||\vk_2|$.
To adjust the correction (\ref{impact-standard}) to the energy scale used in Ref.~\cite{Fadin:2011we}, we need to perform an additional transformation:
\begin{equation}
 \langle G R_1|_{t}\rightarrow \langle G R_1|_{c} = \langle G R_1|_t -\frac12\ln\left(\frac{\qs_2}
   {\qs_1}\right)\langle G R_1|^{(B)}\hat{\cal K}_{m}^{(B)}|{\cal G}_1{\cal G}_2\rangle \,,
 \label{transform to scale}
\end{equation}
where the subscript $c$ means transformed to fit the conformal energy scale and $\hat{\cal K}_{m}^{(B)}$ is the modified LO kernel. Let us put, in a way similar to  Eqs. \eqref{I 1} and  \eqref{I 2},
\begin{equation}
\vec{\Phi}_{GR_1{c}}^{{\cal G}{\cal G}_{2}} \ = \    \vec{\Phi}_{GR_1{t}}^{{\cal G}{\cal G}_{2}}  + \vec{\cal I}_3,\;\;
\label{transformation to scale}
\end{equation}
then the integral for $\vec{\cal I}_3$ can be written as
\begin{equation}
 \vec{\cal I}_3 =-\ln\left(\frac{\qs_2}
{\qs_1}\right)\int\frac{d\vl}{\pi}\left(\vec{C}^{\;\prime}_1-\vec{C}_1\right)\frac{1}{\xps_1\xps_2}  \biggl(\frac{\xs_1\xps_2+\xs_2\xps_1}{\vls} -\qs_2\biggr) \label{I 3}.
\end{equation}
Here instead of $\vec{C}^{\;\prime}_1$ the difference  $\left(\vec{C}^{\;\prime}_1-\vec{C}_1\right)$ is taken  and instead of the full modified kernel only its part related to real gluon production is kept. Moreover, the integral is written in two-dimensional transverse space.
Indeed, due to gluon Reggeization the BFKL kernel for the adjoint representation of the colour group has the eigenvalue which is equal to the gluon trajectory, and the corresponding eigenfunction in the LO is a constant. It means that for the modified kernel the same eigenfunction corresponds to zero eigenvalue. Therefore, in the initial integral with $\vec{C}^{\;\prime}_1$ and the modified kernel   we can change   in the integrand  $\left(\vec{C}^{\;\prime}_1-\vec{C}_1\right)$  with $\vec{C}^{\;\prime}_1$ without change of the integral. After that, the virtual part of the kernel, which conserves Reggeon momenta, can be omitted, and
we come to the integral \eqref{I 3}  which is infrared finite and can be calculated in two-dimensional space.
Integration can be done using the same decomposition  as in Eq.~\eqref{decomposition 1} and the basic integral \eqref{int master}, and we get
\begin{equation}
\vec{\cal I}_3 =-\ln\left(\frac{\qs_2}
{\qs_1}\right)
\Biggl[\vec{C_1}\ln\left(\frac{\q_2^{\;4}(\q_1-\x_1)^4}{\qs_1\xs_2\vks\xs_1}\right)-\vec{C_2}
\ln\biggl(\frac{\qs_2(\q_1-\x_1)^2}{\qs_1\xs_2}\biggr)
\Biggr].
\label{scale transform}
\end{equation}
This result,  together with the transformation (\ref{transformation to scale}) and the correction \eqref{impact-quasi conform} gives
\[
\vec{\Phi}_{G  R_1c}^{{\cal G}_{1}{\cal G}_{2}}=\vec{C}_1\Biggl[
-\ln\left(\frac{\qs_2(\q_1-\x_1)^2}{\xs_1\vks}\right)\ln\left(\frac{\q_2^{\;2}(\q_1-\x_1)^2}{\qs_1\xs_2}\right)
- \ln^2\left(\frac{\qs_1}
  {\qs_2}\right)
-\frac34\ln^2\left(\frac{\vks\xs_1}
 {\qs_1\xs_2}\right)
\]
\[
 -2\frac{(\vks)^\epsilon}{\epsilon^2}+4\zeta(2)\Biggr]
+\frac14\vec{C}_2\ln\biggl(\frac{\qs_2(\q_1-\x_1)^2}{\qs_1\xs_2}\biggr)\ln\biggl(\frac{\q^{\;4}_2(\q_1-\x_1)^4\xs_1\vks}{\x_2^{\;6}\q_1^{\;6}}\biggr)
\]
\begin{equation}
+\frac32 \Bigg[\vec{C}_2\times\Biggl(\Bigl[\x_1\times\x_2\Bigr]I_{\x_1,\x_2}+\Bigl[\q_1\times\x_1\Bigr]I_{\q_1,-\x_1}+\Bigl[\vk\times\x_2\Bigr]I_{\vk,\x_2}+\Bigl[\q_1\times\vk\Bigr]I_{\q_1, -\vk}
\Biggr)   \Biggr] ~. \label{impact-conform}
\end{equation}
This expression gives us the NLO correction to the impact factor for Reggeon-gluon transition  in the scheme with conformal kernel and energy evolution parameter, which were  used for the calculation of the remainder function.  However, it is the impact factor for the full amplitude, not for the remainder function. To obtain the impact factor for the remainder function we have to take  the  impact factor \eqref{GR} with $\Phi_{G  R_1c}^{{\cal G}_{1}{\cal G}_{2}}$ instead of $\Phi_{G  R_1*}^{{\cal G}_{1}{\cal G}_{2}}$  and with the polarisation vector $\ec$ of definite helicity, and to extract from it the piece included in the BDS ansatz.

Let us consider, for definiteness, the production of a gluon with positive helicity, $\ec=(\vec e_x-i\vec e_y)/\sqrt 2$. Then,
\begin{equation}
\ec\vec{C}_1= -\frac{q^-_1r^+_1}{\sqrt 2(q_1-r_1)^+},\;\; ~~~~~~
\ec\vec{C}_2= -\frac{q^-_1q^+_2}{\sqrt 2k^+},
\;\; \frac{\ec\vec{C}_2}{\ec\vec{C}_1}
= 1-z,
\label{C chiral}
\end{equation}
where $z=-q^+_1r^+_2/(k^+r^+_1)$ is the conformal invariant ratio (invariant with respect to M\"{o}bius transformations of complex variables $p_i$ such that $r^+_1=p_1-p_2, r^+_2=p_2-p_3, -q^+_1=p_3-p_4, k^+_1=p_4-p_1$). Chiral components of the vector $\vec{\Phi}_{G  R_1{c}}^{{\cal G}_{1}{\cal G}_{2}}$ \eqref{impact-conform} can be rewritten using the   relations
\begin{equation}
\big[\vec c\times\big[\vec a \times \vec b \bigr]\big]^- =\frac12 c^-[a, b],
\end{equation}
where $[a, b] = a^-b^+-a^+b^-$, and
\[
 \int_0^1\frac{dx}{|x-z|^2}\,
\ln \frac{|z|^2}{x^2}\,
=\,\ \frac{1}{z^+-z^-}\left(2\int_0^1\frac{dx}{x}\,
\ln \frac{1-xz^-}{1-xz^+} -\ln|z|^2\ln \frac{1-z^-}{1-z^+} \right)
\]
\begin{equation}
=\,\ \frac{1}{z^+-z^-}\left(-2\int_0^1\frac{dx}{x}\,
\ln \frac{1-x/z^-}{1-x/z^+} -\ln|z|^2\ln \frac{(1-z^-)z^+}{(1-z^+)z^-} \right).
\end{equation}
Taking into account these relations and Eq.~\eqref{I p q 1} we have
\[
[\vec c\times [\vec a\times \vec b]]^-
I_{\vec a, \vec b} = \frac{c^-}{2}\left(2\int_0^1\frac{dx}{x}\,
\ln \frac{(1+xa^-/b^-)}{(1+xa^+/b^+)} -\ln \bigl(\frac{\vec a^{\; 2}}{\vec b^{\; 2}}\bigr)\ln \frac{(a+b)^- b^{+}}{(a+b)^+ b^{-}} \right)
\]
\begin{equation}
=\frac{c^-}{2}\left(-2\int_0^1\frac{dx}{x}\,
\ln \frac{(1+xb^-/a^-)}{(1+xb^+/a^+)} -\ln \bigl(\frac{\vec a^{\; 2}}{\vec b^{\; 2}}\bigr)\ln \frac{(a+b)^-a^+}{(a+b)^+a^-} \right)~.
\label{a b c I chiral}
\end{equation}
Then, we transform the sum of dilogarithms which are  obtained from the correction~\eqref{impact-conform} with the help of the relation \eqref{a b c I chiral} using the identity
\begin{equation}
Li_2(-\frac{b}{a})+Li_2(-\frac{c}{a})
+Li_2(-\frac{b}{d})+Li_2(-\frac{c}{d})=Li_2(\frac{bc}{ad})-\frac12
\ln^2\left(\frac{a}{d}\right),
\end{equation}
where  $a+b+c+d=0$.  As result,
after some algebra and with account of Eq. \eqref{GR} we obtain
\[
\langle G  R_1|{\cal G}_{1}{\cal G}_{2}\rangle \ = \  \langle G  R_1|{\cal G}_{1}{\cal G}_{2}\rangle^{(B)}\, \Biggl\{1+\frac{\bar g^2}{8} \Biggl[(1-z)\Biggl(
\ln\biggl(\frac{|1-z|^2}{|z|^2}\biggr)\ln\biggl(\frac{|1-z|^4}{|z|^6}\biggr)
\]
\[
-6Li_2(z)+6Li_2(z^*) -3\ln|z|^2
\ln\frac{1-z}{1-z^*} \Biggr)-4\ln|1-z|^2\ln\frac{|1-z|^2}{|z|^2} -3\ln^2|z|^2
\]
\begin{equation}
 -4 \ln^2\left(\frac{\qs_1}
  {\qs_2}\right)
 -8\frac{(\vks)^\epsilon}{\epsilon^2}+16\zeta(2)\Biggr]\Biggr\}~. \label{impact-almost conform}
\end{equation}
Finally, in order to move to the impact factor for the calculation of the  remainder function, one has to
discard the terms $\bar g^2\left(-(1/2) \ln^2\left({\qs_1}/{\qs_2}\right)-{(\vks)^\epsilon}/{\epsilon^2}+2\zeta(2)\right)$ in the impact factor (\ref{impact-almost conform}), since they are already taken into account in the BDS ansatz.


\section{Conclusion}
In this paper, we have calculated in the next-to-leading order  the impact factor for Reggeon-gluon transition in the  maximally extended supersymmetric Yang-Mills theory (N=4 SYM) with large number of colours. Our final goal was the impact factor for the calculation of the high energy behaviour of the remainder function for the BDS ansatz. On the way to this goal we have obtained several noteworthy intermediate results.

In the next-to-leading order impact factors are scheme dependent. First, we have found the impact factor in the bootstrap scheme, which was used in Refs.~\cite{Kozlov:2011zza}-\cite{Kozlov:2013zza} for the check of validity of the bootstrap condition, the last and the most complicate in the set of the  conditions, the fulfillment of which provides the multi-Regge form of production amplitudes.  Starting from rather cumbersome  results of
Refs.~\cite{Kozlov:2011zza}-\cite{Kozlov:2013zza} for Yang-Mills theories with  any  number of fermions and scalars in arbitrary representations of the  gauge group, after great simplifications we have obtained a simple expression  for  the impact factor in the bootstrap scheme for $N=4$ SYM with large number of colours. Then, we have transformed it in the standard scheme. To reach our goal, we needed  to have the impact factor in the scheme with  conformal invariant kernel and energy evolution parameter (M\"{o}bius scheme). The impact factor in the M\"{o}bius scheme was obtained by the transformation from the standard scheme. Finally,  the impact factor for the calculation of the remainder function was obtained from the impact factor in the M\"{o}bius scheme by subtraction of the terms contained in the BDS ansatz. It turns out that this impact factor is invariant with respect to M\"{o}bius transformations in  momentum space.  Definitely, it is the reaffirmation of justice of the conjecture  about dual conformal invariance of the remainder function.
From the other side, it can be considered as a cross-check of a large number of calculations in the BFKL theory.

\vspace{0.5cm} {\textbf{{\Large Acknowledgments}}}

\vspace{0.5cm} V.S.F. thanks the Dipartimento di Fisica
dell'Universit\`{a} della Calabria and the Istituto Nazionale di
Fisica Nucleare (INFN), Gruppo Collegato di Cosenza, for warm hospitality
while part of this work was done and for financial support.

\end{document}